\documentstyle[12pt]{article}
\begin{document}
\baselineskip 0.75cm
\begin{center}
\Large{\bf{Tight-binding study of the electronic and magnetic 
properties of an L1$_0$ ordered FeCu alloy}}

\vspace{1.25cm}
\large{Daniel Errandonea\footnote{e-mail address: 
daniel@ges1.fisapl.uv.es

Fax number: 34 - 6 - 398 - 3146}

\vspace{0.5cm}
Departamento de F\'{\i}sica Aplicada, Universidad de Valencia, 

C/ Dr. Moliner 50, E-46100 Burjasot (Valencia), Spain}
\end{center}

\vspace{2.5cm}
\begin{abstract}
We have calculated the electronic structure of the tetragonal L1$_0$
ordered FeCu by solving self-consistently a tight-binding Hamiltonian 
for s, p and d electrons. We have found by total energy calculation 
that this structure is ferromagnetic. In addition, we have determined 
that the equilibrium ratio between the interlayer and the intralayer 
lattice parameters is 0.947.
\end{abstract}

\vspace{0.5cm}
PACS numbers: 75.50.Rr, 75.50.Bb, 73.20.At

\vspace{5cm}

\pagebreak

Magnetic multilayers sinthesized by ultrahigh vacuum deposition
techniques have received much attention recently for both the
fundamental and technological points of view\cite{1,2}. Particularly,
the fabrication of ordered alloys with layered structure has created
widespread interest because they may have important application 
potential in magnetic recording\cite{3}. A typical example is the 
tetragonal L1$_0$ ordered structure shown in Fig. 1, which consists of 
alternate atomic monolayers of two different elements. The magnetic 
properties of artificially fabricated Fe/Pt and Fe/Au multilayers with 
L1$_0$ structure has been experimentally investigated\cite{4}. 
Moreover, theoretical calculations of the magnetic and structural 
properties of Fe/Au monoatomic multilayers have been made by means of 
the FLAPW method\cite{5}. All these systems have been found to be 
ferromagnetic (FM) and to have a large uniaxial magnetic anisotropy 
perpendicular to the atomic layers. Another system which promises 
interesting magnetic properties is the tetragonal L1$_0$ ordered FeCu. 
The fabrication of this alloy has not been reported yet, but it 
seems possible since it is widely believed that fcc Fe ($\gamma$-Fe) 
films grow coherently on the Cu(100) surface due to the small lattice 
mistmatch (1.1 $\%$). In fact, layer-by-layer growth has been reported 
for Fe/Cu(100) films grown by thermal deposition\cite{memmel} and has 
been greatly improved very recently by pulsed laser 
deposition\cite{jenniches}. In these systems, the layers of Fe are 
found to be individually FM, but regarding to the interlayer coupling, 
ferromagnetism has been observed only for the first layers, whereas 
the deeper ones exhibit antiferromagnetism\cite{kraft,muller}. 

Motivated by these results, we have undertaken, for the first time
to the best of our knowledge, a theoretical investigation of the
electronic structure of the tetragonal L1$_0$ ordered FeCu. 
Our calculations have been performed using the self-consistent 
tight-binding (TB) method within the unrestricted Hartree-Fock (HF) 
approximation by considering s,p and d orbitals. We have chosen this 
approach because it has shown to give succesful results in several 
magnetic systems as Fe clusters\cite{6} and Fe monolayers sandwiched
in noble metals\cite{7}. In addition, this formalism has the advantage 
that it is posible to increase the complexity of the model by adding 
one by one different contributions to the Hamiltonian. Whitin this 
simple model, we have found, by total energy calculations, that the 
L1$_0$ ordered FeCu alloy is FM with a magnetic moment of 2.69 $\mu_B$ 
and that the ratio of the interlayer lattice constant ($c$) to the 
intralayer lattice constant ($a$) is 0.947. As there are neither 
experimental data nor theoretical results available in the literature 
over the artificial Fe/Cu L1$_0$ structure, we limit ourselves to 
compare the present results with the related Fe/Au L1$_0$ structure. 

The spin-polarized electron structure of the system is determined by 
solving self-consistently the TB Hamiltonian including hopping to
second nearest neigbohrs. The Hamiltonian in the HF approximation has 
the general form:
\begin{eqnarray}
H=\sum_{\vec{i},m,\sigma}E_{\vec{i}m\sigma}n_{\vec{i}m\sigma} +
\sum_{\vec{i}\not=\vec{j},m,m^{'},\sigma} T_{\vec{i}\vec{j}}^{mm^{'}}
a_{\vec{i}m\sigma}^{\dag}a_{\vec{j}m^{'}\sigma} \quad ,
\end{eqnarray}

\noindent
where $a_{\vec{i}m\sigma}^{\dag}$ ($a_{\vec{i}m\sigma}$) and
$n_{\vec{i}m\sigma}$ are the creation (annihilation) and
number operators of an electron state at atomic site $\vec{i}$ in 
the orbital $m$ with spin $\sigma$. $E_{\vec{i}m\sigma}$ and 
$T_{\vec{i}\vec{j}}^{mm^{'}}$ stand for the single HF energies
and the hopping integrals. The value of these integrals, for Fe-Fe
and Cu-Cu interactions, are taken from Ref. \cite{8} and for Fe-Cu
interactions we employ the geometrical averages of the preceding
ones.

The electron-electron interaction and the magnetic effects are 
introduced only in the diagonal terms of the Hamiltonian through
the on-site potencial shift of the energy levels,
\begin{eqnarray}
\Delta E_{\vec{i}m\sigma} = \sum_{m^{'}} U_{\vec{i}mm^{'}} \Delta 
n_{\vec{i}m^{'}} - \frac{\sigma}{2} J_{\vec{i}m} \mu_{\vec{i}m} \quad,
\end{eqnarray}

\noindent
where $U_{\vec{i}mm^{'}}$ and $J_{\vec{i}m}$ are the intra-atomic 
Coulomb and exchange integrals. To reduce the number of parameters
we have assumed $U_{\vec{i}ss}=U_{\vec{i}sp}=U_{\vec{i}pp}$ and
$U_{\vec{i}sd}=U_{\vec{i}pd}$; thus, we have only three kinds of
intra-atomic Coulomb integrals. The value of these parameters
for Fe and Cu were estimated as in the work by Sarma\cite{10}.
The exchange integrals were neglected with the exception of 
the one corresponding to Fe d electrons, whose value 
($J_{Fe}$ = 1.19 eV) was chosen to recover the proper
magnetic moment of bulk fcc Fe ($\mu_{Fe}$ = 2.5 $\mu_B$).
Regarding to $\Delta n_{\vec{i}m}$ and $\mu_{\vec{i}m}$,
$\Delta n_{\vec{i}m} = n_{\vec{i}m} - n_{\vec{i}m}^0$
is the difference between the number of electrons determined by 
$n_{\vec{i}m} = <n_{\vec{i}m\uparrow}> + <n_{\vec{i}m\downarrow}>$
and the occupation in the bulk metallic configuration 
$n_{\vec{i}m}^0$, and $\mu_{\vec{i}m} = <n_{\vec{i}m\uparrow}> -
<n_{\vec{i}m\downarrow}>$ is the local magnetic moment. Both 
magnitudes are determined from $<n_{\vec{i}m\sigma}>$, which is
calculated self-consistently through:
\begin{eqnarray}
<n_{\vec{i}m\sigma}> = \int_{-\infty}^{E_F}
\rho_{\vec{i}m\sigma} (E) \delta E \quad,
\end{eqnarray}

\noindent
where the Fermi level ($E_F$) is determined from the global 
charge neutrality condition, and the spin-polarized local density
of states (SLDOS) $\rho_{\vec{i}m\sigma} (E)$ is calculated by the
recursion method\cite{11}. The number of k levels of the continuous
fraction expansion (k=30) is chosen so that $\rho_{\vec{i}m\sigma} 
(E)$ becomes independient of k.

Sistematically we have calculated the electron structure of the
present system as a function of the volume. Intralayer ferromagnetism
with both FM and antiferromagnetic (AFM) coupling between Fe layers
were considered and $a$ = $a_{Cu}$ = 3.61 $\AA$ was assumed. By 
calculating the total energy, taking care of the double counting of 
the Coulomb and magnetic interactions, we determine the stability of 
the two different phases and the optimum ratio $c/a$. The results are
plotted in Fig. 2. There it can be seen that the FM phase is more
stable because is lower in energy than the AFM phase. By fitting the
total energy through a Murnaghan equation of state\cite{mur},
we determine for the FM case an equilibrium ratio $c/a$ = 0.947, 
which is very close to the value estimated from the hard sphere radii 
of Cu and Fe ($c/a$ = 0.974)\cite{radii}. This optimum ratio 
corresponds to a contraction of about 5 $\%$ with respect to the Cu 
volumen and results to be larger than the value obtained in total 
energy calculations in the Fe/Au L1$_0$ structure\cite{5}.  

Next we focus on magnetization. For $c/a$ = 0.947 we have found that 
Fe has a magnetic moment ($\mu_{Fe}$) of 2.69 $\mu_B$. In addition, we 
have obtained that the atoms of Cu have a small magnetic moment 
($\mu_{Cu}$). Both moments are plotted as a function of $c/a$ in Fig. 
3. It can be seen there that $\mu_{Fe}$ increases from 2.49 $\mu_B$ 
to 2.81 $\mu_B$ and $\mu_{Cu}$ decreases from 0.112 $\mu_B$ to 
0.038 $\mu_B$ as $c/a$ variates from 0.8 to 1. This is just the
behaviour that one would expect, because in general an enhancement of
the magnetic moment occurs when diluiting the Fe atoms and no magnetic
moment would be expected in isolated Cu atoms. 

To understand the enhancement of $\mu_{Fe}$ 
and the origin of $\mu_{Cu}$ it is a good help to give a look
to the SLDOS. Fig. 4 shows the local density of states corresponding to
the FM case for the equilibrium c/a value. The spin-up peaks around -2 
eV and spin-down peaks around 0 eV are mainly d electrons of Fe. The 
increase of the spliting between spin-up and spin-down peaks in 
comparison with the bulk spliting explains 
the enlargement of the Fe magnetic moment. Regarding to the
peaks around -4 eV, they are related to d electrons of Cu atoms.
It can be seen in Fig. 4 that they are a little bit shifted from
each other giving rise to the small magnetic moment of Cu atoms.
The existence of such small induced magnetic moment in Cu is 
related with the coupling between Fe monolayers and Cu monolayers.

Comparing again with the Fe/Au L1$_0$ structure, it results that
in the present system $\mu_{Fe}$ is less enhanced. We think that this 
difference may be due to two facts: the first one is that the indirect 
exchange interaction between Fe layers across noble metal layers
decreases monotonically along the series Cu $\rightarrow$ Ag 
$\rightarrow$ Au \cite{12}. The second one is that, d electrons of Cu 
are much closer in energy from those of Fe than d electrons of Au 
(situated around -5 eV\cite{5}) and, as a consequence of the larger 
overlap of d electrons, the hybridization between Fe and Cu becomes 
larger than the hybridization between Fe and Au. This also allows
us to understand why $\mu_{Cu} > \mu_{Au}$. Then, as we have obtained, 
$\mu_{Fe}$ must be less enhanced in the Fe/Cu L1$_0$ structure than in 
the Fe/Au L1$_0$ structure because of the smaller degree of 
confinement experienced by Fe monolayers in the former system due to 
the larger indirect coupling and hybridization strength. We think that 
another possible consequence of this facts would be the decrease of 
the strong magnetic anisotropy observed in Fe/Au structures\cite{4}. 
In any case, only comparison with already non-existent experimental 
work would allow us to check our predictions.

In summary, we have investigated theoretically the electronic and 
magnetic properties of the Fe/Cu monoatomic multilayer by means of a
tight-binding scheme. We have found that in the tetragonal L1$_0$
structure the FM coupling between Fe layers is more favorable than 
the AFM one. For the FM case we determine that the ratio $c/a$ has a 
value of 0.947 and the magnetic moment of Fe atoms is 2.69 $\mu_B$. 
From our results, we conclude that Fe/Cu L1$_0$ structure belongs to 
the tetragonal L1$_0$ family of ferromagnets. Nevertheless, further 
progress is strongly needed for accurate understanding of its 
properties.

\pagebreak

\pagebreak

{\Large {\bf Figure captions}}

\vspace{1cm}
{\bf Figure 1:}Crystal structure of the tetragonal L1$_0$ ordered 
structure.

\vspace{0.5cm}
{\bf Figure 2:}Relative total energy as a function of the volume
for the tetragonal L1$_0$ ordered FeCu. Circles are the calculated 
values when ferromagnetic ($\bullet$) or antiferromagnetic ($\circ$) 
coupling is considered. Lines are least square fits to the Murnaghan
equation of state. V$_0$ is the equilibrium volume of fcc Cu.

\vspace{0.5cm}
{\bf Figure 3:}Calculated magnetic moment of Fe ($\bullet$) and Cu 
($\circ$) in the tetragonal L1$_0$ ordered FeCu as a function of 
$c/a$. Lines are only a guide to the eye.

\vspace{0.5cm}
{\bf Figure 4:}Local density of states (SLDOS) for the tetragonal
L1$_0$ ordered FeCu. The vertical solid line represents the Fermi
level.


\begin{thebibliography}{20}

{\small
\bibitem{1} Farrow R.F.C., Parkin S.S.P., Dobson P.J., Neave J.H.
and Arrot A.S. {\it Thin Films Growth Techniques for Low-dimensional
Structures} (Plenum, New York, 1987).

\bibitem{2}Schuller I.K. {\it Physics Fabrication and 
Applications of Multilayered Structures} edited by Dhez D. and 
Weisbuch C. (Plenum, New York, 1988) p. 139.

\bibitem{3}Klemmer T., Heydick D., Okimura H., Zhang B. and Soffa W.A.
{\it Scr. Metall. Mater.} {\bf 33} (1995) 1793.

\bibitem{4}Takanashi T., Mitani S., Sano M. and Fujimori H.
{\it J. Appl. Phys.} {\bf 67} (1995) 1016 and references therein.

\bibitem{5}Shi Z., Cooke J.F., Zhung Z. and Klein B. 
{\it Phys.Rev.} {\bf B 54} (1996) 303.

\bibitem{memmel}Memmel N. and Detzel T. {\it Surf.Sci.} {\bf 307-309}
(1994) 490.

\bibitem{jenniches}Jenniches H., Klaua M., H\"oche H. and
Kirschner J. {\it Appl. Phys. Lett.} {\bf 69} (1996) 3339.

\bibitem{kraft}Kraft T., Marcus P. and Scheffer M. {\it Phys.Rev.}
{\bf B 49} (1994) 11511.

\bibitem{muller}M\"uller S. {\it et al} {\it Phys.Rev.Lett.} {\bf 74}
(1995) 765.

\bibitem{6}Bouarab S., Vega A., Alonso J.A. and I\~niguez M.P.
{\it Phys.Rev.} {\bf B 54} (1996) 3003.

\bibitem{7}Errandonea D. {\it J.Phys.:Condens.Matter} {\bf 7} (1995)
9439.

\bibitem{8}Andersen O.K. {\it Highlights of Condensed Matter
Theory} edited by Bassani F., Fumi F. and Tossi M. 
(North Holland, Amsterdam, 1985).

\bibitem{10}Sarma D.D. and Bandyopadhyay T. {\it Phys.Rev.} {\bf B 39} 
(1989) 3517.

\bibitem{11}Haydock R. {\it Solid States Physics} {\bf 35}
edited by Ehrenreich E., Seitz F. and Turnbull D. 
(Academic Press, London, 1980) p.215.

\bibitem{mur}Murnaghan F.D. {\it Proc. Natl. Acad. Sci.} {\bf 30}
(1944) 244.

\bibitem{radii}$c$ = 2 $\sqrt{(r_{Fe} + r_{Cu})^2 - (a/2)}$, 
where $r_{Fe}$ and $r_{Cu}$ are the hard sphere radii of Fe and Cu, 
respectively.

\bibitem{12}Bruno P. and Chappert C. {\it Phys.Rev. Lett.}
{\bf 67} (1991) 1602.
}
\end{thebibliography}
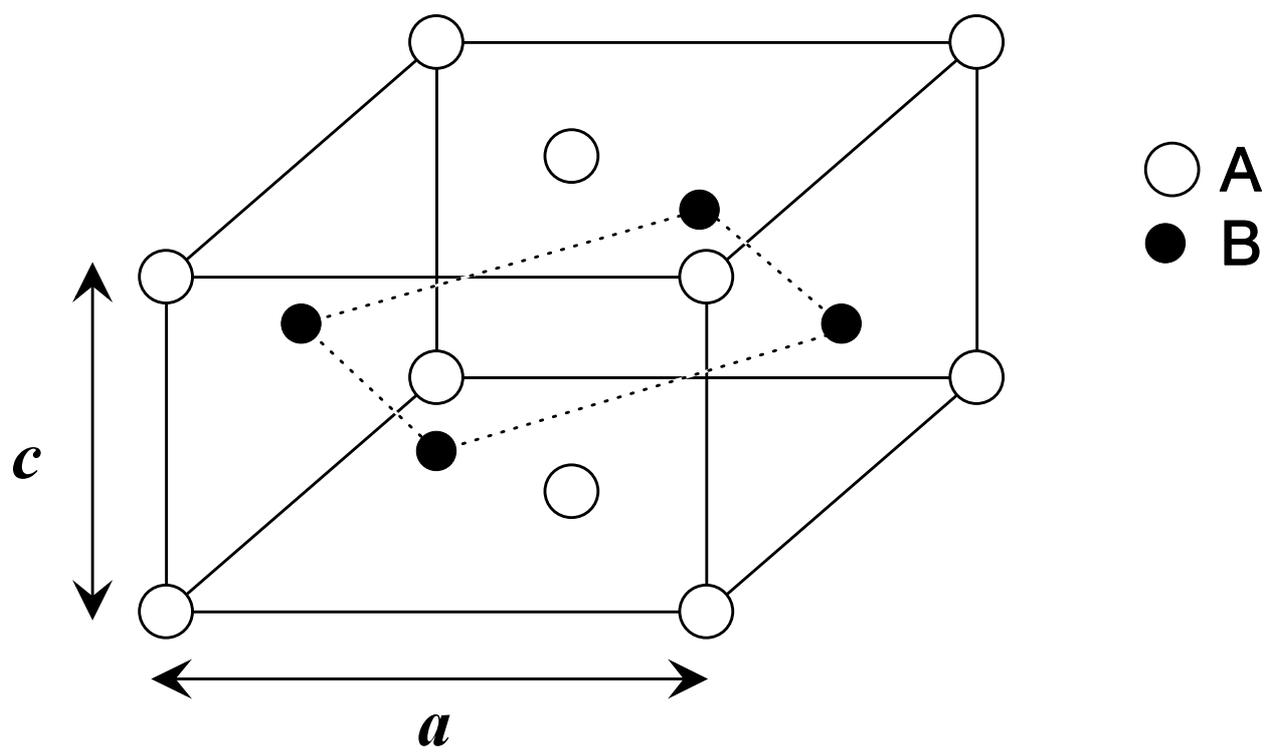
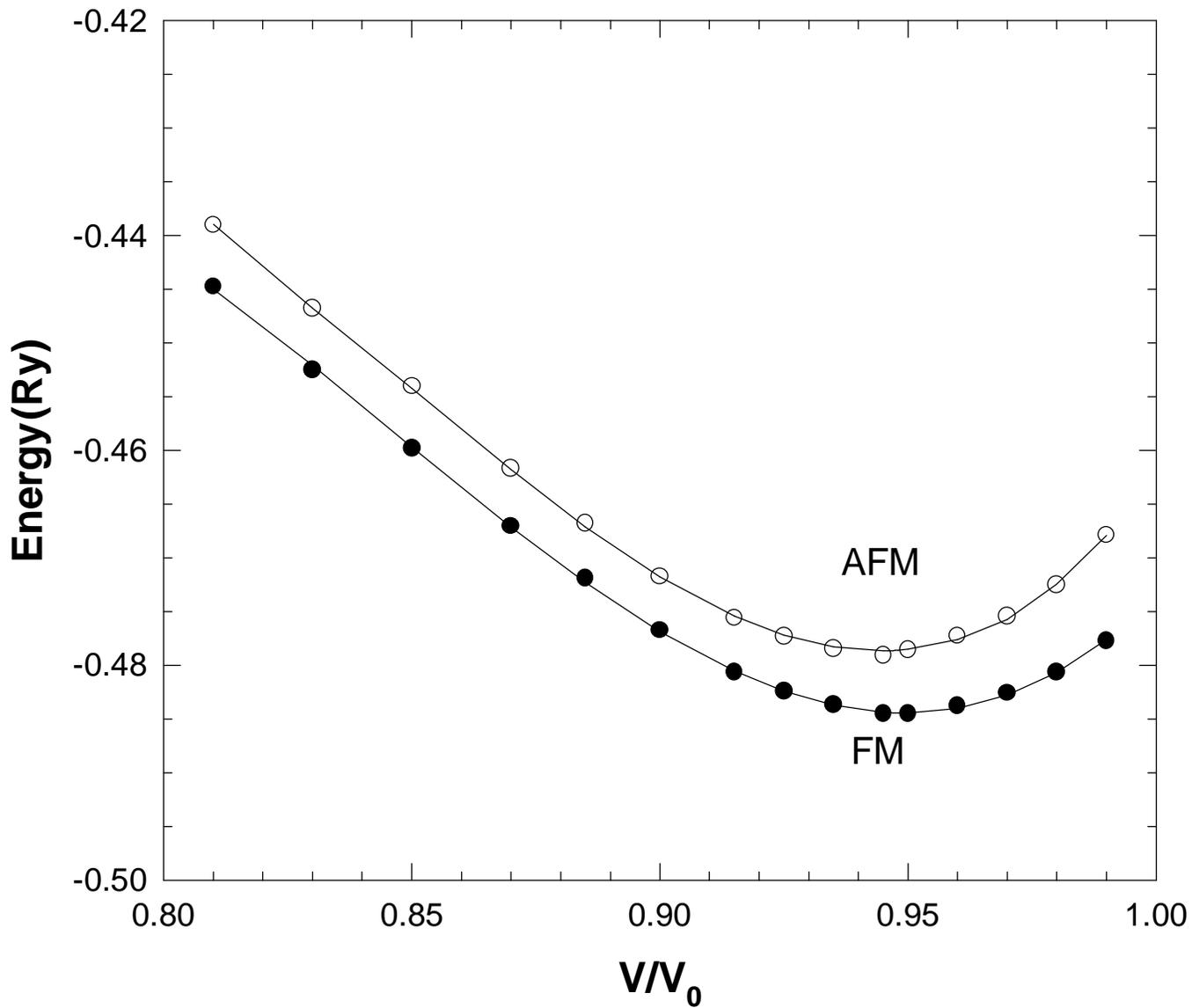
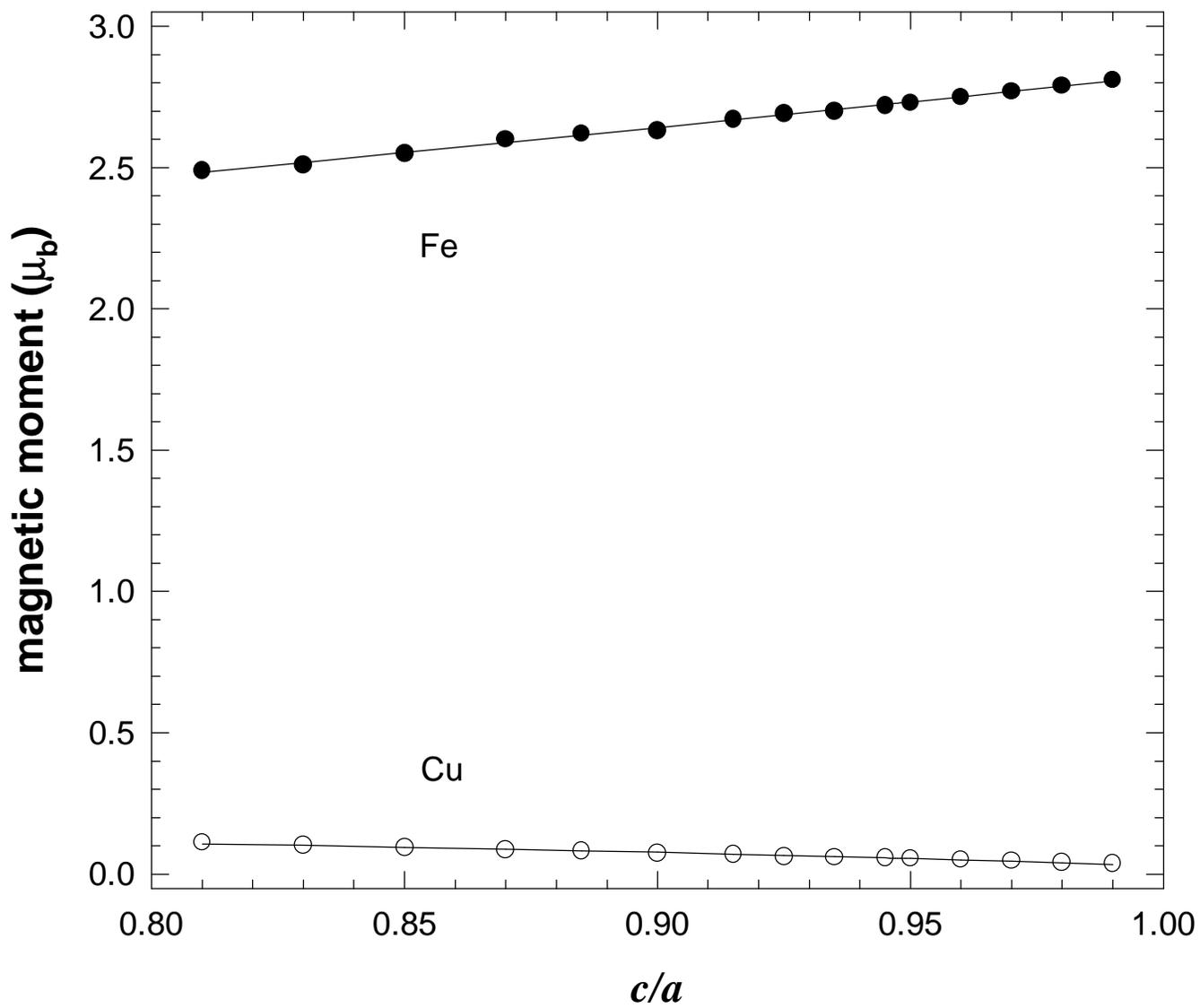
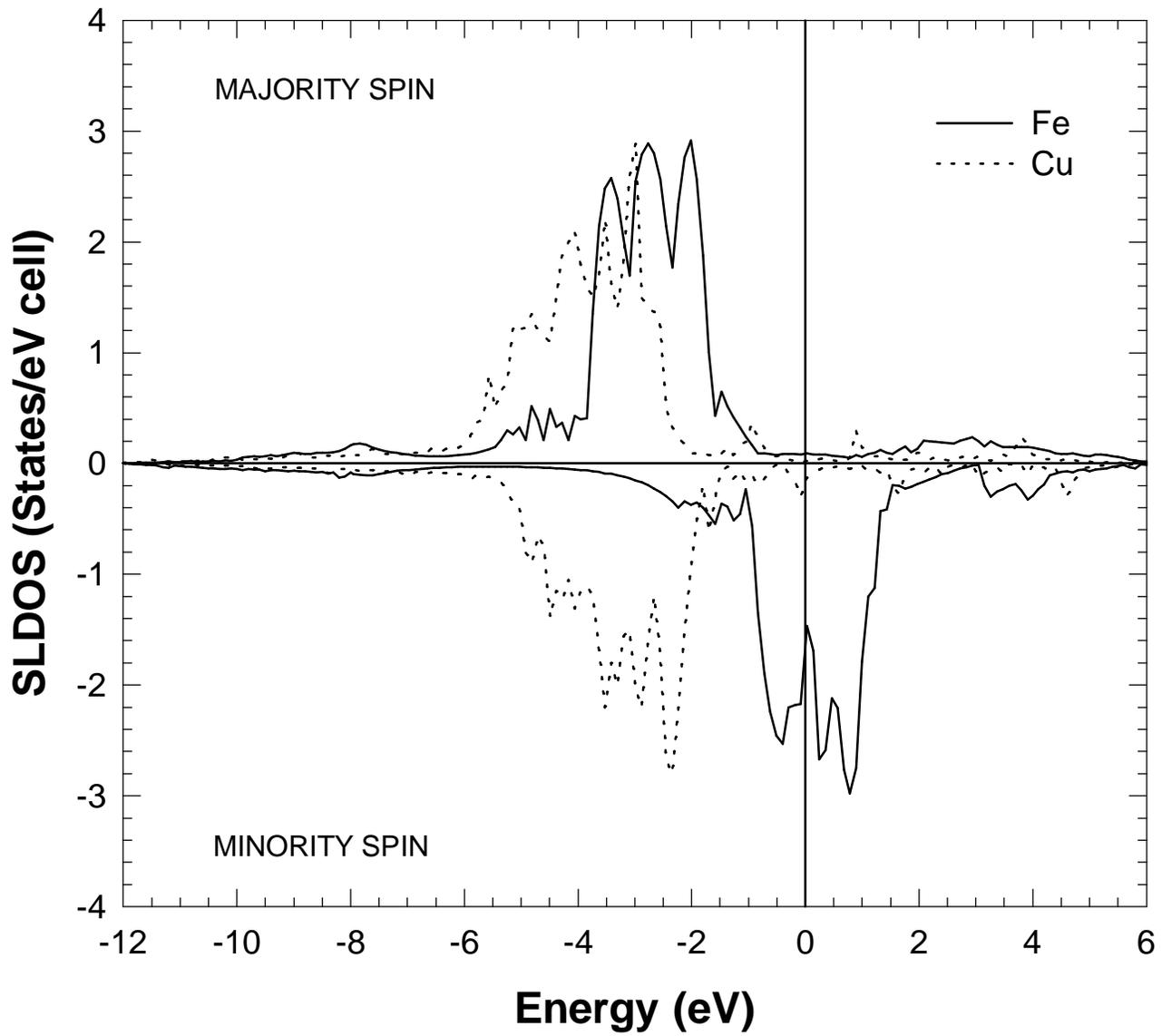
\end{document}